\documentclass[a4paper,11pt]{article}
\usepackage{pos}

\title{FIMP Dark Matter at the LHC}

\author[a,b]{Susanne Westhoff\,}

\affiliation[a]{Institute for Mathematics, Astrophysics and Particle Physics, Radboud University, 6500 GL Nijmegen, The Netherlands}
\affiliation[b]{Nikhef, Science Park 105, 1098 XG Amsterdam, The Netherlands}

\emailAdd{susanne.westhoff@ru.nl}

\abstract{This brief summary targets feebly interacting massive particles, FIMPs, which are interesting candidates for dark matter. The cosmic history of FIMP dark matter often leads to predictions of long-lived mediator particles at laboratory experiments. I point out the role of the LHC in searching for such particles and sketch its complementarity with low-energy experiments.}

\FullConference{The Eleventh Annual Conference on Large Hadron Collider Physics (LHCP2023)\\
 22-26 May 2023\\
 Belgrade, Serbia\\}

\begin{document}
\maketitle

\section{Introduction}
\noindent Dark matter, if it has a particle nature, is likely to interact only little with visible matter, that is, particles of the Standard Model (SM). Null results in laboratory searches for dark matter particles, as well as astrophysical observations, suggest that any possible dark sector must be well hidden from current observation. Small dark matter interactions are not only phenomenologically motivated. They are also predicted in models that explain the dark matter abundance today from dynamical processes in the early universe.

Such dark matter candidates are often referred to as FIMPs, Feebly Interacting Massive Particles, in analogy with their weakly interacting, but heavier predecessors in research, WIMPs. By convention, the term FIMP typically refers to GeV-scale particles whose fundamental interactions are tiny, typically weaker than the Standard Model forces at observable energies. The cosmic history and search strategy for FIMP dark matter can be drastically different from the methods developed for WIMPs. The goal of this brief summary is to sketch the current state-of-the-art of FIMP dark matter, with a focus on the interplay between early universe dynamics and collider searches.

In Sec.~\ref{sec:cosmic}, I review possible cosmic histories for FIMP dark matter and how they predict a dark sector including long-lived particles (LLPs). In Sec.~\ref{sec:LHC}, I give representative examples of searches for long-lived particles at the LHC, focusing on axion-like particles. Finally, in Sec.~\ref{sec:conclusions}, I explain what we can learn from such searches about FIMP dark matter.

\section{Cosmic history of a FIMP}
\label{sec:cosmic}
\begin{figure}[t!]
    \centering
   \includegraphics[width=1.0\textwidth]{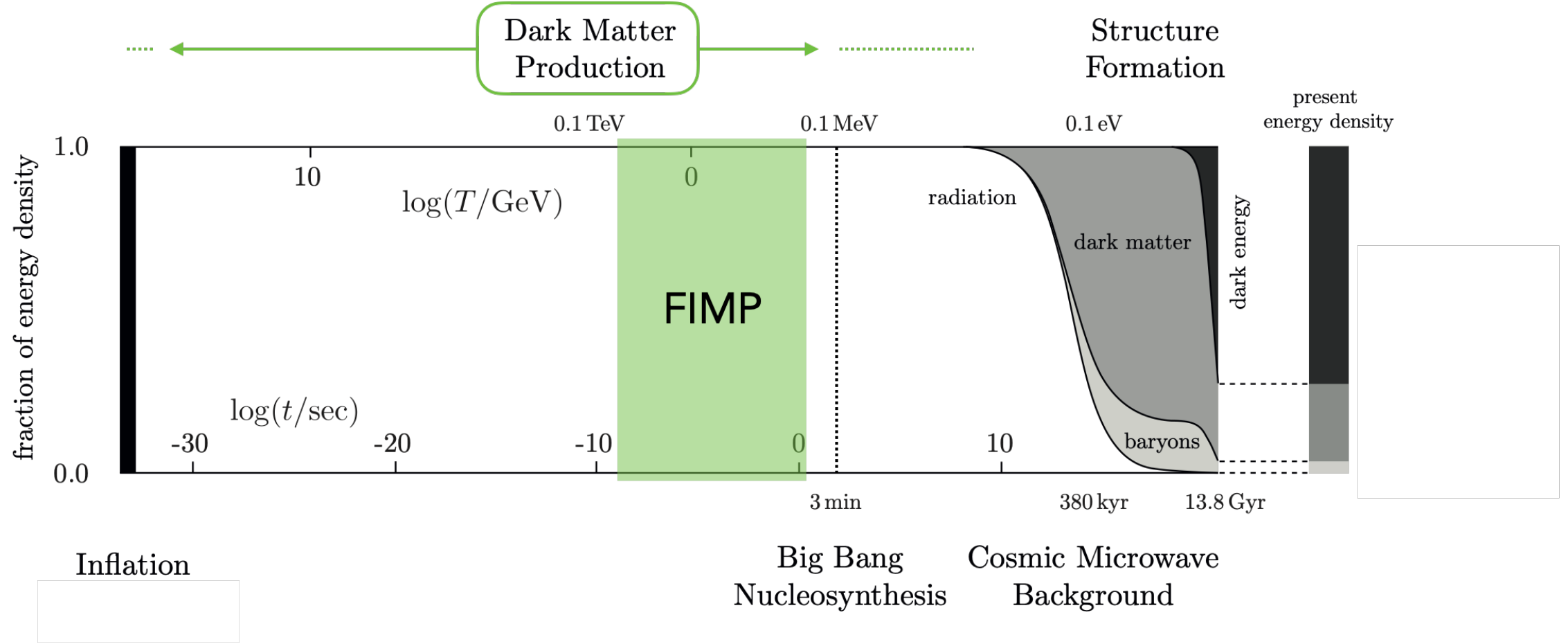}
    \caption{Feebly interacting massive particles (FIMPs) in the early universe. Adapted from Daniel Baumann.\label{fig:fimps}}
\end{figure}

\noindent Any dark matter candidate should fit into the evolution of particle species in the early universe, in order to explain the large-scale structures in the universe and the temperature fluctuations we observe in the cosmic microwave background. Fig.~\ref{fig:fimps} shows the evolution of the energy densities in the universe in time and temperature.

In principle, dark matter could have been produced at any time before structure formation. In many scenarios, the dark matter abundance we observe today can be related to a dynamical process. For freeze-out from thermal equilibrium, this process is tied to the mass scale of the dark matter particles. Freeze-in, on the other hand, is usually most effective around the mass scale of a dark partner which feeds the dark matter density through decays or annihilation.

For FIMPs with GeV-scale masses, freeze-out happens not long before big bang nucleosynthesis. Freeze-in can occur long before that time; typically at energy scales well above the TeV scale. In either case, viable scenarios of FIMP dark matter should not leave any imprints on the measurements of the abundance of light elements. Dark sector particles other than dark matter itself should therefore have decayed before the time of nucleosynthesis.

FIMP dark matter, $\chi$, usually freezes out through processes other than pair annihilation into SM particles, for instance through co-annihilation with a dark partner, $\eta$, via $\chi \eta \to SM$, or through pair annihilation into dark partners via $\chi\chi \to \eta\eta$, known as secluded dark matter~\cite{Pospelov:2007mp}. Both processes are relevant if pair annihilation via $\chi\chi \to SM$ is suppressed, i.e., if dark matter is only feebly interacting with the SM particles. For co-annihilation to be efficient, the relative mass splitting between dark matter particle and the dark partner has to be small, $\Delta m \ll 1$.

For freeze-in~\cite{Hall:2009bx}, dark matter must be initially out of equilibrium. This requires even smaller couplings to SM particles than in the freeze-out scenarios described above.

Phenomenologically, the most important common feature in these scenarios is the presence of a dark partner with suppressed couplings to the Standard Model. This implies suppressed decays $\eta\to SM$, or $\eta \to \chi SM$ in the case of co-annihilation. In the latter case, the decay is also phase-space suppressed by the small mass splitting $\Delta m$. As a result, the dark partners have macroscopic lifetimes at the scales of colliders. For freeze-in, the lifetime can exceed collider scales and the dark partners appear as essentially stable.

This relation between early-universe dynamics and collider signatures with long-lived particles is a highly non-trivial prediction. It suggests that FIMP dark matter scenarios can be probed at colliders with searches for displaced or invisible decays of dark partners.

\section{FIMPs at the LHC}
\label{sec:LHC}
\begin{figure}[t!]
    \centering
   \includegraphics[width=0.55\textwidth]{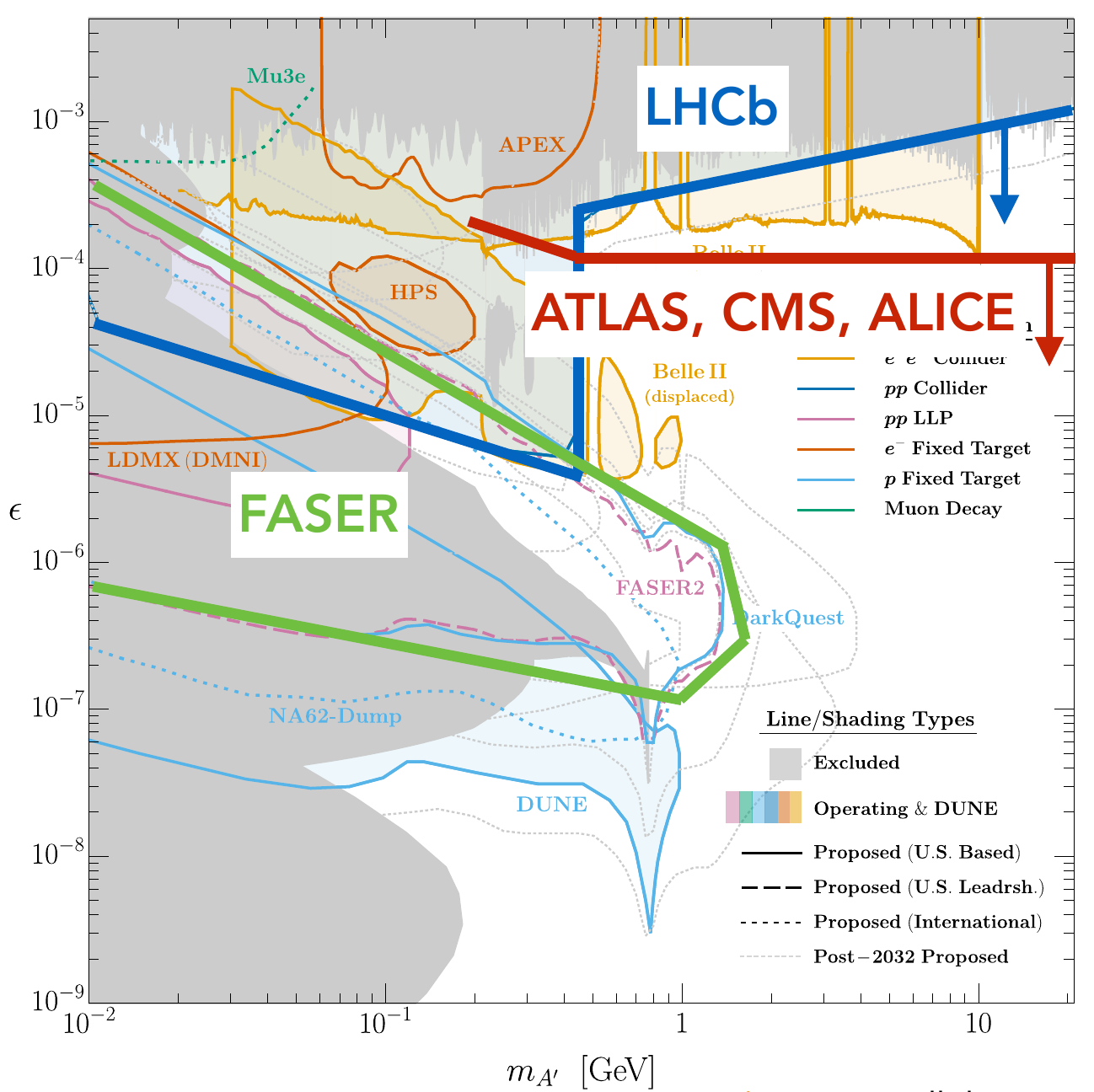}
    \caption{Searches for dark mediators at colliders and fixed-target experiments. Here: dark photon searches as a function of the mass $m_{A'}$ and the kinetic mixing parameter $\epsilon$. Adapted from Ref.~\cite{Batell:2022dpx}.\label{fig:searches}}
\end{figure}

\noindent Searches for long-lived particles cover a wide range of experiments and signatures. For GeV-scale particles, the main players are collider experiments (ATLAS, CMS, LHCb, Belle II), and long-baseline experiments (fixed-target, beam dump). Fig.~\ref{fig:searches} shows a summary of dark photon searches as an example of probing one-particle extensions of the Standard Model. Most searches rely on displaced vertex signatures.

The LHC plays an important role in these searches at several fronts: At high energies, it is the only running experiment that can probe long-lived particles with masses above $10\,$GeV. At small couplings, the new LHC-based experiment FASER competes with other long-baseline experiments for the highest sensitivity to sub-GeV particles with very long decay lengths.

In what follows, I demonstrate the complementarity of the different experiments and signatures at the example of axion-like particles (ALPs). In general, ALPs are pseudo-scalar particles originating from a spontaneously broken global symmetry. In the context of FIMP dark matter, ALPs can be the carriers of a new force between dark matter and the Standard Model. Searching for long-lived ALPs then means searching for feebly interacting dark force mediators.

\paragraph{LHCb and Belle II} At flavor experiments, ALPs $a$ can be efficiently produced in meson decays, mostly via $B \to K^{(\ast)} a$~\cite{Batell:2009jf}. ALPs with masses below the hadronic threshold (three pions) decay into lepton pairs via $a\to \mu^+\mu^-$, $a\to e^+e^-$, or into photons via $a\to \gamma\gamma$. For small couplings, the decay appears as a displaced vertex or as missing energy if it happens outside the detector.

\begin{figure}[t!]
    \centering
   \includegraphics[width=0.6\textwidth]{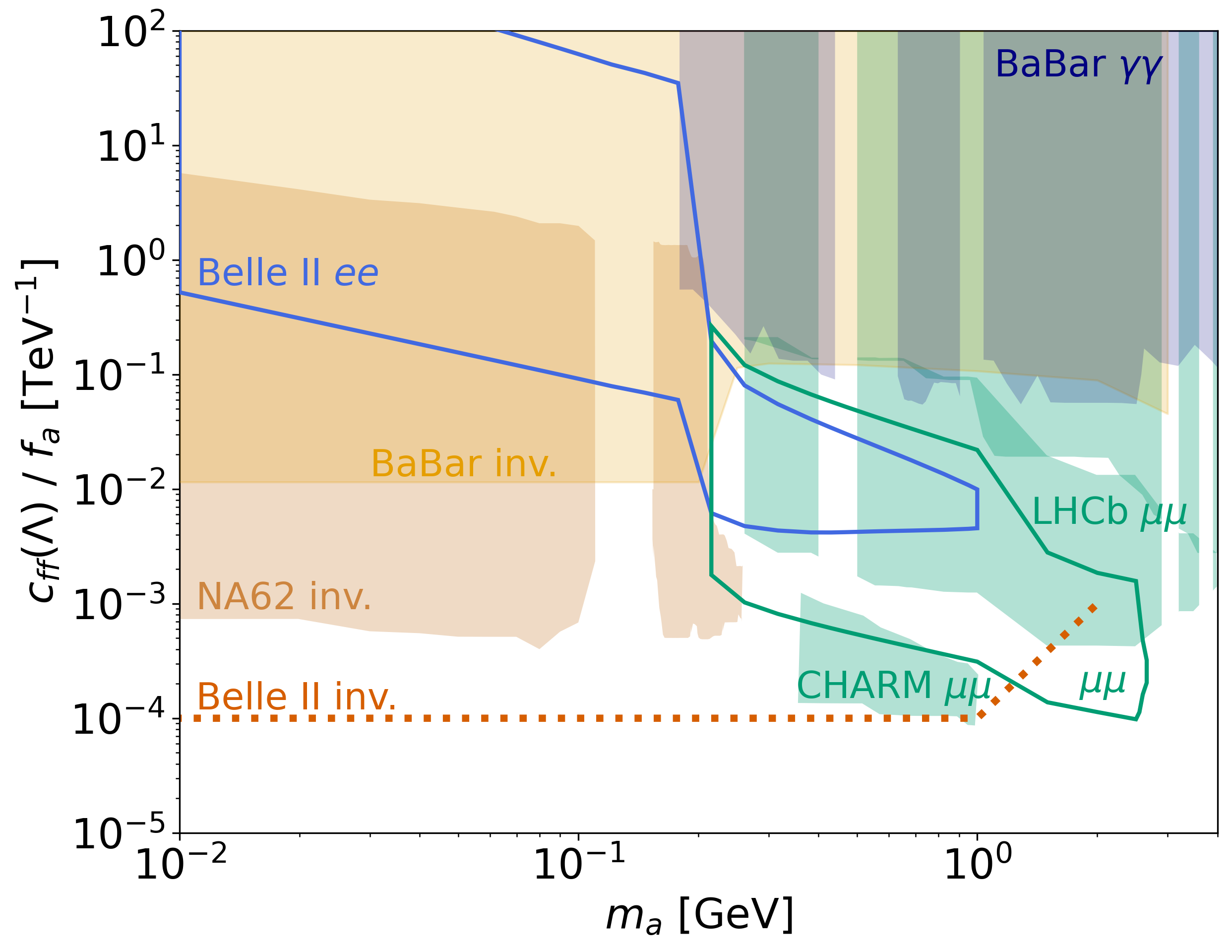}
    \caption{Bounds on ALPs from flavor observables~\cite{Ferber:2022rsf}. Shown are existing bounds on the flavor-universal ALP coupling to fermions, $c_{ff}$, as a function of the ALP mass, $m_a$, from various searches for ALPs in meson decays at flavor and fixed-target experiments. The dashed line shows the expected sensitivity of Belle II in a search for $B\to K a, a\to \text{inv.}$\label{fig:alp-bounds-flavor}}
\end{figure}

Fig.~\ref{fig:alp-bounds-flavor} shows the complementarity of searches for ALP signatures with displaced vertices ($ee,\mu\mu,\gamma\gamma$) and with missing energy (``inv.''). ALPs with masses above the di-muon threshold decay mostly within the detector. Lighter ALPs have such a large decay length that missing energy searches perform best. Even lighter feebly coupling ALPs can be probed with kaon decays $K\to \pi a$ at fixed-target experiments.

\begin{figure}[t!]
    \centering
   \includegraphics[width=0.7\textwidth]{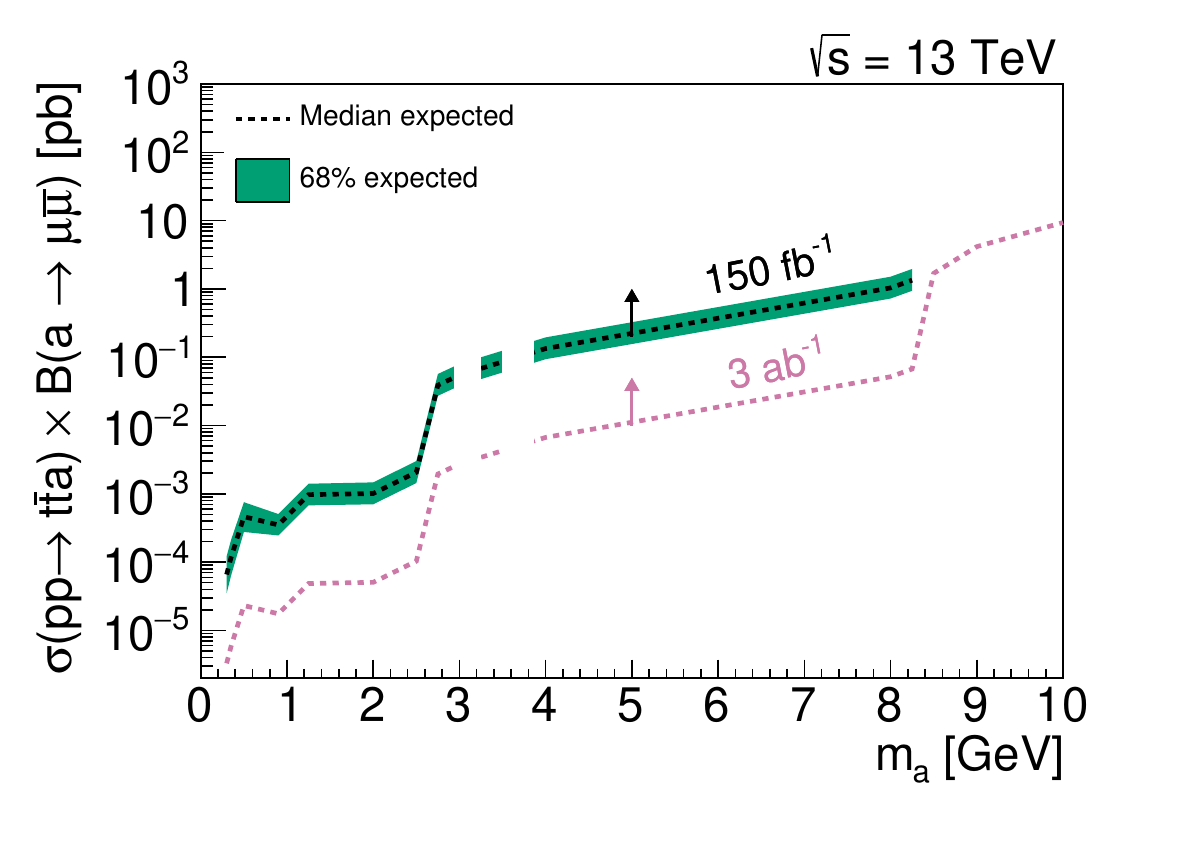}
   \vspace*{-0.5cm}
    \caption{Sensitivity predictions for ALPs produced in association with a top-antitop pair and decaying to a displaced di-muon vertex, $pp\to t\bar{t}a, a\to\mu^+\mu^-$~\cite{Rygaard:2023dlx}. Shown are the expected 95\% C.L. upper bounds on the ALP coupling to top quarks, $c_{tt}/f_a$, as a function of the ALP mass, $m_a$, as they could be obtained at the LHC with Run III data (black dashed) and at the HL-LHC (rose dashed).\label{fig:alp-bounds-displaced}}
\end{figure}

\paragraph{ATLAS and CMS} The LHC detectors in the central phase-space region offer to probe long-lived ALPs through a large variety of signatures. Much explored are Higgs decays, which are very sensitive to ALP couplings to the Higgs boson, but also to the top quark and weak gauge bosons through loop-induced decays~\cite{Bauer:2017ris}. Alternatively, ALPs can be produced from resonant top quarks, for instance in association with a top-antitop pair~\cite{Esser:2023fdo,Rygaard:2023dlx,Blasi:2023hvb,Phan:2023dqw}. For long decay lengths, a displaced di-lepton vertex in association with a top-antitop pair is a promising signature~\cite{Rygaard:2023dlx}. In particular, the presence of the top quarks offers a well-known object to trigger on. Fig.~\ref{fig:alp-bounds-displaced} shows the expected sensitivity of the LHC to such an ALP signature during Run III and with future data collection at the HL-LHC. The search has the potential to probe ALPs with masses up to about $8\,$GeV, where hadronic decays decrease the lifetime and render the final-state reconstruction complicated. For very light ALPs below the di-electron threshold, the decay length can be so large that top-associated missing energy searches apply~\cite{Esser:2023fdo}, with a higher sensitivity than displaced vertices. The situation is thus comparable to meson decays: missing energy searches lead for light, long-lived ALPs; displaced vertex searches for heavier ALPs with shorter lifetimes.

\paragraph{FASER} The new cool kid in town is a long-baseline experiment with the LHC as source, exploring events produced in the forward direction at the ATLAS interaction point. In such events, light, long-lived particles can be produced from meson decays or bremsstrahlung and detected $500\,$m further down the line in the FASER detector. Recently, FASER has published its first search for long-lived dark photons, which also probes a freeze-out dark matter scenario~\cite{CERN-FASER-CONF-2023-001}. This is an interesting first demonstration that searches for light dark matter mediators can be conducted at the LHC in the forward region, in an environment that differs from that of fixed-target experiments.

\section{Conclusions}
\label{sec:conclusions}
\noindent What do we learn from long-lived particle searches about FIMP dark matter? Oftentimes, there is no 1:1 connection between the cosmic history of a FIMP and its collider phenomenology. A useful way to classify viable dark matter scenarios in terms of the coupling constants of the mediator to dark matter and SM particles are so-called \emph{mesa diagrams}~\cite{Chu:2011be,Bharucha:2022lty}. Such diagrams illustrate that the parameter space of one model comprises several possible cosmic histories.

For ALPs as mediators, several recent analyses point out FIMP scenarios from freeze-in or freeze-out variations~\cite{Bharucha:2022lty,Fitzpatrick:2023xks,Dror:2023fyd,Armando:2023zwz}. To pin down the cosmic evolution and fundamental properties of the FIMP, broader analyses are needed that combine observations from astrophysics and laboratory experiments.

As shown in this brief write-up, the LHC plays an important role in searching for light, long-lived mediators. This might appear counter-intuitive, given that the LHC was designed as a machine to explore heavy new particles. With its high sensitivity at the lifetime frontier, the potential to discover light, feebly coupling new particles cannot be underestimated.

\bibliographystyle{JHEP}
\bibliography{skeleton}

\end{document}